\documentstyle[12pt,psfig]{article}
\begin{document}

{\flushright{\small NSF-ITP/97-036\\
astro-ph/9704263\\}}

\vspace{0.2in}
\begin{center}
{\Large  {\bf Is there evidence for cosmic anisotropy}\\
{\bf in the polarization of distant radio sources?}}

\vspace{0.2in}
Sean M.  Carroll
\vskip 0.2cm
{{\it Institute for Theoretical Physics, University of
California, \\
Santa Barbara, California 93106, USA}
\\
\small E-mail: {\tt carroll@itp.ucsb.edu}}

\vspace{0.2in}
George B. Field
\vskip 0.2cm
{{\it Harvard-Smithsonian Center for Astrophysics \\
60 Garden St., Cambridge, Massachusetts 02138, USA}
\\
\small E-mail: {\tt field@cfa.harvard.edu}}
\end{center}

\vskip 1truecm

\begin{abstract}
Measurements of the polarization angle and orientation of
cosmological radio sources may be used to search for unusual
effects in the propagation of light through the universe.  Recently,
Nodland and Ralston have claimed to find evidence for a redshift-
and direction-dependent rotation effect
in existing data.  We re-examine these data and argue that there
is no statistically significant signal present.  We are able to place
stringent limits on hypothetical chiral interactions of photons 
propagating through spacetime.
\end{abstract}

\medskip

\newpage

\renewcommand{\baselinestretch}{1}
\baselineskip 18pt

\section{Introduction}
\label{sec:introduction}

The polarization of radiation emitted by
distant radio galaxies and quasars offers a
way to search for chiral effects in the propagation
of electromagnetic radiation.  Such objects are often elongated in
one direction, so that one may define a position angle $\psi$ which
describes the orientation of the object on the sky.  Synchrotron
radiation can lead to a significant linear polarization of the source, 
and the angle $\chi$ of the plane of
polarization may also be measured \cite{ss}.  
(The angle of polarization will typically undergo Faraday rotation, but this 
effect can be removed by using the fact that Faraday rotation is 
proportional to the square of the wavelength.)  One can therefore
study the relative angle $\chi-\psi$ between the position and
polarization vectors, keeping in mind that this quantity is only defined 
modulo $180^\circ$.  It has been found \cite{cks, hc} that $\chi-\psi$
is not distributed randomly; there is a large peak at $\chi-\psi \approx
90^\circ$, and a smaller enhancement at $\chi-\psi \approx 0^\circ$.
Since many of these sources are at significant redshifts, and therefore
very far away, testing whether this relationship is maintained for
distant sources provides constraints on possible chiral
effects on the propagation of light through the universe, which could
rotate $\chi-\psi$ away from the intrinsic value (that which would
be measured at the source).

From a field theory point of view, the simplest such chiral effect
arises from a Lagrange density
\begin{equation} 
  {\cal L} = -{{1}\over 4} F_{\mu\nu} F^{\mu\nu} - \phi
  F_{\mu\nu} \widetilde{F}^{\mu\nu}\ .
  \label{lag}
\end{equation}
Here, $F_{\mu\nu} = \partial_\mu A_\nu - \partial_\nu A_\mu$ is the field 
strength for the electromagnetic field $A_\mu$, $\widetilde{F}^{\mu\nu}
=(1/2)\epsilon^{\mu\nu\rho\sigma}F_{\rho\sigma}$ is the dual field
strength, and $\phi$ is a pseudoscalar field which need not be fundamental
(it can be a function of other fields in the theory).  We set $c=1$
throughout.  This Lagrangian is the simplest way to couple a neutral
pseudoscalar to electromagnetism in a parity-invariant way, and often
describes the effective coupling of pseudoscalar particles (such as
pions or axions) to photons.

Our interest here is in the case where $\phi$ varies only very slowly 
over extremely large distances.  In that case an electromagnetic wave
travelling through the background $\phi$ field will undergo a rotation
in its polarization state which depends on the change in $\phi$; such
an effect arises in a variety of contexts \cite{sikivie}-\cite{hk}.
In the WKB limit where the length scale for variations in $\phi$ is
much larger than the wavelength of the photon, the polarization
angle $\chi$ obeys the simple relation
\begin{equation} 
  \Delta\chi = \Delta\phi\ ,
%  \label{}
\end{equation}
where $\Delta$ indicates the change between source and observer.
(Here, and in Eq.~(\ref{deltachi}) below, $\Delta\chi$ is measured
in radians; elsewhere we measure all angles in degrees;
no confusion should arise.)  This effect is independent of
wavelength, and can therefore be distinguished from ordinary
Faraday rotation.
Carroll, Field and Jackiw \cite{cfj} suggested that observations of
polarized radio sources provide a stringent test of such an effect,
since they afford an opportunity to constrain
$\Delta\phi$ over a large interval in space and time (see also
\cite{cdff,gt}).

The specific model investigated in \cite{cfj} set $\partial_\mu\phi
= -(1/2)p_\mu$, where $p_\mu$ is a 4-vector whose expectation
value parameterizes violation of Lorentz invariance (as well as
CPT \cite{cgl}).  It was
hypothesized that there exists a preferred coordinate frame,
close to the background Robertson-Walker frame of our universe, in
which $\partial_\mu p_\nu=0$.  This implies that the predicted
rotation of the polarization angle for a source at redshift $z$
is given in terms of the timelike component $p_0$ and the
spacelike vector $\vec p$ by
\begin{equation} 
  \Delta\chi = - {1\over 2} r(p_0 - p \cos\theta)\ ,
  \label{deltachi}
\end{equation}
where $p=|\vec p \,|=(\delta^{ij}p_ip_j)^{1/2}$, $\theta$ is the angle 
between $\vec p$ and the direction
toward the source, and $r$ is the proper spacelike distance 
traveled.  If we take a flat ($k=1$) universe
as a reasonable approximation, we have
\begin{equation} 
  r = {2\over{3H_0}}\left[1-(1+z)^{-3/2}\right]\ ,
%  \label{}
\end{equation}
where $H_0$ is the Hubble constant today.  Regardless of whether or
not one is interested in tests of Lorentz invariance, Eq.~(\ref{deltachi})
is a useful parameterization of potentially observable chiral
effects.

In \cite{cfj} it was shown that the radio galaxies at redshift greater
than 0.4, with maximum polarizations greater than $5\%$, were strongly
clustered around $\chi-\psi\approx 90^\circ$, using a sample of
galaxies and redshifts obtained from the literature \cite{cks,hc,bc,sdma}.
Assuming that the timelike component $p_0$ would
be significantly larger than the spacelike part $\vec p$, the limit
\begin{equation} 
  p_0 \leq 1.7\times 10^{-42} h_0~{\rm GeV}
  \label{limit}
\end{equation}
was obtained, where $h_0 = H_0/100$~km/sec/Mpc.  Recently, Nodland
and Ralston \cite{nr}, using the same set of data,\footnote{As mentioned
in \cite{nr}, there were a handful of transcription errors in the table
published in \cite{cfj}.  The corrected values are: entry 9, coordinates
0106+72; entry 35, coordinates 0459+25; entry 84, $\psi=39^\circ$;
entry 124, coordinates 1626+27; entry 144, $z=0.054$; and entry 153,
$z=0.0244$.  We are grateful to Borge Nodland for informing us of
these corrections.} searched for anisotropic effects such as those
that would arise from a nonzero spacelike part $\vec p$ in
Eq.~(\ref{deltachi}).  Surprisingly,
they claimed to find a significant signal in the data.  Given the
fundamental importance of such a result, we have undertaken a
re-examination of the data, and present our results in this paper.
We conclude that the data are most consistent with
no effect, contrary to \cite{nr}.  Our disagreement stems primarily
from the method used to disentangle the $180^\circ$ ambiguity in
the quantity $\chi-\psi$, and the use of randomly generated data
for comparison purposes, as will be shown below.

As this manuscript was being completed we received a preprint
by Eisenstein and Bunn \cite{eb}, who come to conclusions similar 
to those expressed in this paper.

\section{Data}
\label{sec:data}

We consider the same set of data as was used in \cite{cfj}, with
the corrections noted above.  This set includes 160 sources,
with redshifts as high as 2.012.  The first step is to establish
the existence of a reliable correlation between $\chi$ and $\psi$,
so that we may test its behavior for distant galaxies.  Following
\cite{nr}, we have divided the data into distant and nearby
objects, with the division drawn at $z=0.3$; there are 89 sources
with $z<0.3$, and 71 with $z\geq 0.3$.  In Figures One and Two, we
have plotted histograms of $\chi-\psi$ for these two sets.  We have 
for the purposes of these figures
defined $\chi-\psi$ so that it lies between $0^\circ$ and $180^\circ$,
and grouped the data into bins which are $10^\circ$ wide.
\begin{figure}
  \vskip -0.75cm
  \centerline{
  \psfig{figure=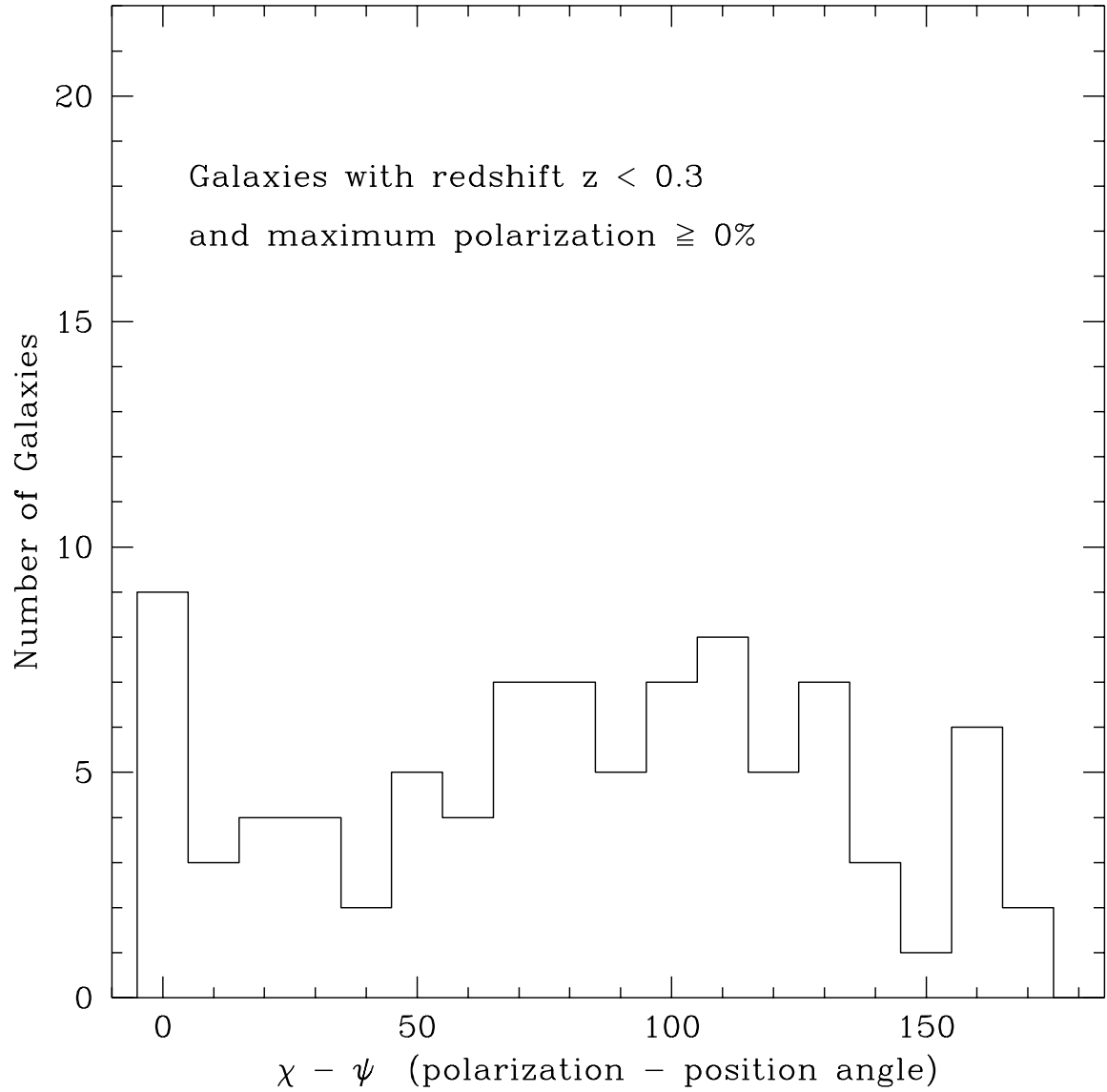,angle=0,height=16.5cm}}
  \vskip -0.1cm
  \caption{Histogram of number of galaxies vs. $\chi-\psi$, for
  galaxies with $z<0.3$.}
\end{figure}

\begin{figure}
  \vskip  -0.75cm
  \centerline{
  \psfig{figure=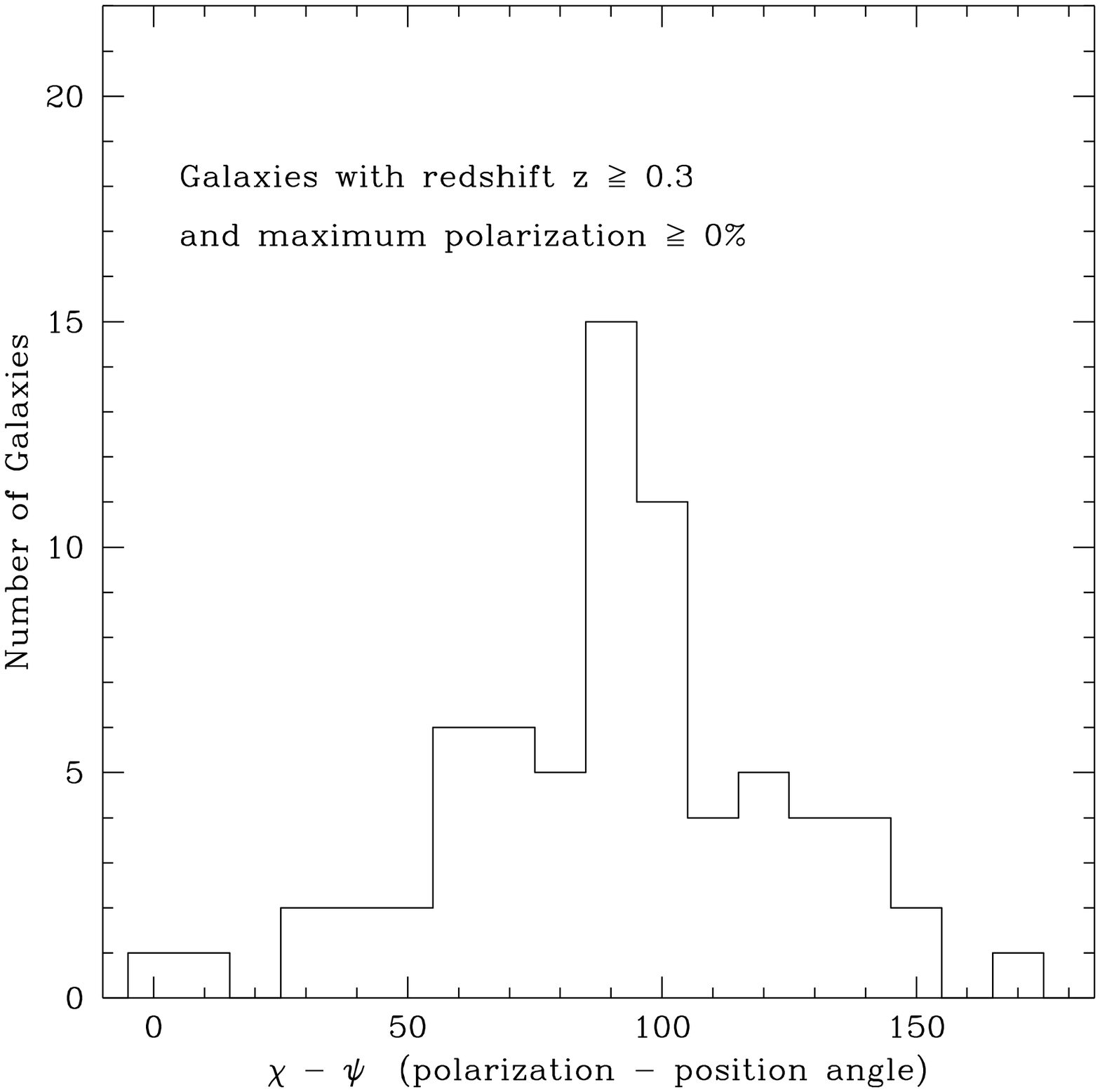,angle=0,height=16.5cm}}
  \vskip  -0.1cm
  \caption{Histogram of number of galaxies vs. $\chi-\psi$, for
  galaxies with $z\geq0.3$.}
\end{figure}

In Figure One, representing nearby galaxies, there appears to be
evidence for a narrow enhancement at $\chi-\psi\approx 0^\circ$, and
a broader peak at $\chi-\psi\approx 90^\circ$.  However, the
correlation is evidently not very strong.  The most likely explanation
for this fact is that many of these galaxies are members of a 
different population than those at high redshift, which
can be observed only if they are of high luminosity,
and the lower-luminosity galaxies demonstrate a weaker correlation
between polarization and position angle.  For the more distant
galaxies shown in Figure Two, there is a very clear peak at
$\chi-\psi\approx 90^\circ$.  There is no noticeable peak at
$\chi-\psi\approx 0^\circ$ in this sample; again, this may be 
explained if the galaxies with $\chi-\psi\approx 0^\circ$ are
members of a lower-luminosity population.  (See \cite{cks,hc} for
discussion of these correlations and their interpretation in
terms of models of the sources.)

It is reasonable to suppose that the galaxies with a higher degree
of maximum polarization would show any effect more strongly than
those that are polarized only weakly.  We therefore show in Figures
Three and Four the same plots as in Figures One and Two, this time
limited only to those sources with maximum polarization greater than
or equal to $5\%$.  These are the sources that were analyzed in
\cite{cfj}.
\begin{figure}
  \vskip -0.75cm
  \centerline{
  \psfig{figure=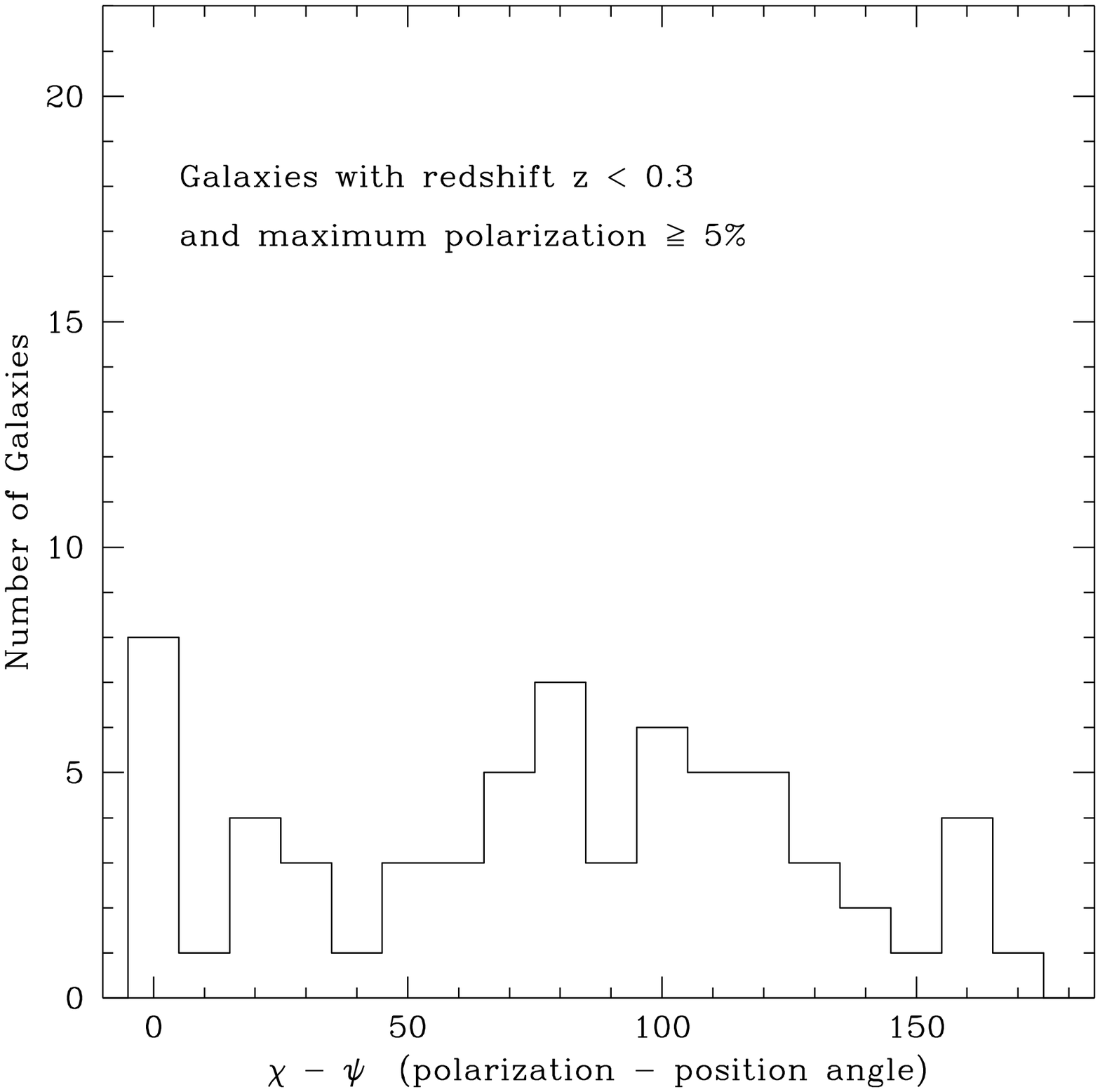,angle=0,height=16.5cm}}
  \vskip -0.1cm
  \caption{Histogram of number of galaxies vs. $\chi-\psi$, for
  galaxies with $z<0.3$ and maximum polarization $\geq 5\%$.}
\end{figure}

\begin{figure}
  \vskip -0.75cm
  \centerline{
  \psfig{figure=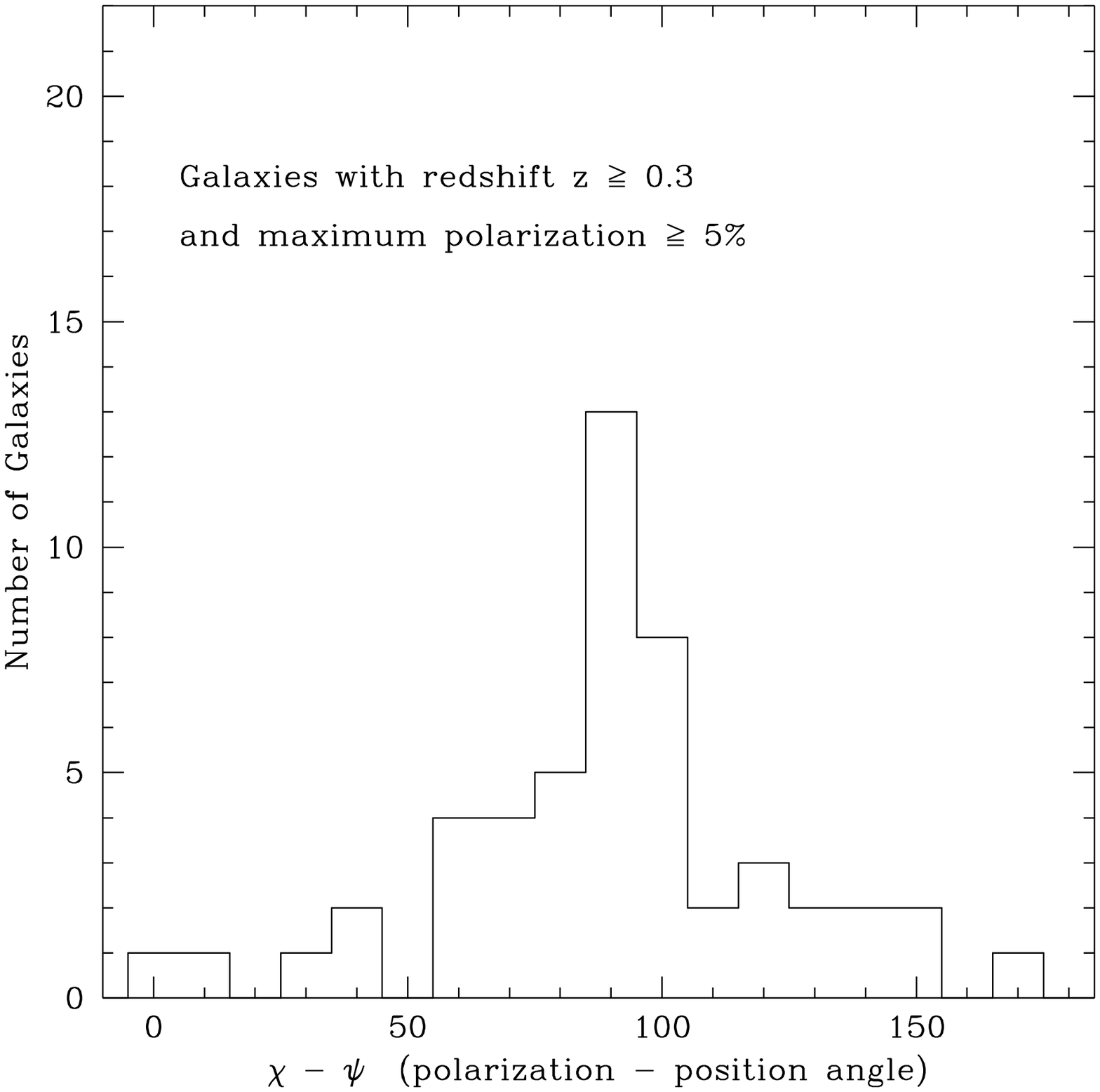,angle=0,height=16.5cm}}
  \vskip -0.1cm
  \caption{Histogram of number of galaxies vs. $\chi-\psi$, for
  galaxies with $z\geq0.3$ and maximum polarization $\geq 5\%$.}
\end{figure}

The peaks found in the previous plots, at $0^\circ$ and $90^\circ$
at low redshifts and more dramatically at  $90^\circ$ in the
high-redshift sample, are also
evident in Figure Three and Figure Four, arguably more convincingly.
However, any increased correlation is offset to some degree by the
smaller number of data points.  Therefore, in the remainder of this
paper we will not discard those sources with maximum polarization
$<5\%$; we will thus use precisely the same data as were analyzed
in \cite{nr}.

The crux of our disagreement with \cite{nr} can be found in Figure
Two, the distribution of $\chi-\psi$ for sources with $z\geq 0.3$.
As noted in \cite{cfj}, this plot constitutes vivid evidence that
the polarization angle in these sources is intrinsically perpendicular
to the position angle on the sky.  If the claim of \cite{nr} is true,
it is necessary to believe that the peak at $90^\circ$ is an accident,
and these data are actually drawn from a distribution which is 
intrinsically centered at $0^\circ$, with position- and redshift-dependent
contributions of order $180^\circ$.  We will argue in the next section
that this is not the simplest interpretation of these data.

\section{Constraints on chiral effects}
\label{sec:constraints}

Searching for a signal in the polarization data is complicated by
the fact that $\chi-\psi$ is only defined modulo $180^\circ$.
In testing any specific hypothesis, it is necessary to choose some
reasonable procedure for resolving this ambiguity.  The method
chosen by the authors of \cite{nr} was the following:
for any choice of direction for the vector
$\vec p$, define an angle $\beta=\chi-\psi\pm 180^\circ$ 
which is between $0^\circ$ and $180^\circ$ if $\cos\theta \geq 0$
and between $-180^\circ$ and $0^\circ$ if $\cos\theta < 0$, where
$\theta$ (which they called $\gamma$) is the angle between $\vec p$
and the direction toward the source.

It was noted in \cite{nr} that this procedure necessarily introduces
correlations between $\beta$ and $r\cos\theta$.  It would be
illegitimate, therefore, to take a statistical correlation between
these two quantities as itself evidence of a signal in the data.
However, if the degree of correlation were much higher than that which
would be expected if there were no signal in the data, we might
conclude that there was a measurable effect.

It is at this point in the analysis that we find two important flaws in the
procedure followed in \cite{nr}.  First, one must reliably determine
the zero point for $\chi-\psi$, which would be observed in the absence
of any chiral effects.  In \cite{nr}, the authors searched for a best
fit to the data of the form $\beta = (1/2)\Lambda_s^{-1} r\cos\theta
+\delta$, where in the notation of Eq.~\ref{deltachi}, $\Lambda_s = p^{-1}$.
They found that the favored value for the zero point was $\delta\approx
0^\circ$.  This seems to be inconsistent with the evidence of
Figure Two, which exhibits a peak at $90^\circ$.  The resolution is
simply the fact that the definition of $\beta$, as described above,
separates the data into two groups, one with $-180^\circ<\beta<0^\circ$
and one with $0^\circ<\beta<180^\circ$.  With this procedure the
favored value for $\delta$ will always be near $0^\circ$; it arises
essentially from taking the average of a group of points clustered
around $90^\circ$ and another clustered around $-90^\circ$.  This 
method of resolving the $180^\circ$ ambiguity is therefore inappropriate
for data which lie naturally in the vicinity of $90^\circ$.

Nevertheless, \cite{nr} argues that the correlation found is
statistically significant, as it was only very rarely reproduced in
artificially generated sets of data.  The procedure for generating
these sets is the second important flaw that we find.  Figure Two
provides evidence that, regardless of the position of the source on
the sky, $\chi-\psi$ is distributed approximately in a Gaussian
distribution centered on $90^\circ$; a best fit to the Gaussian
yields a dispersion of $\sigma = 33^\circ$.  Therefore, in searching
for position-dependent effects, it is appropriate to compare the
actual data to data which is generated by drawing from a similar
distribution.  In \cite{nr}, on the other hand, artificial realizations
were generated completely randomly, {\it i.e.} from a flat probability
distribution for $\chi-\psi$.  This has a dramatic effect on the
claimed significance of the result.  We performed an independent
analysis,\footnote{There are two procedures described in \cite{nr}.
In the first, the correlation in the artificial data sets was calculated
using the best-fit direction obtained from the real data.
As the hypothesis being tested is that there is some
direction of anisotropy, not that a specific direction is picked out
{\it a priori}, it is more appropriate to compare correlations in
each set of data (real and artificial) calculated using the best-fit direction
for that set.  This is equivalent to the second procedure in \cite{nr}, 
and is the method followed here.} using two
different methods of generating the artificial data sets: first by
drawing from a flat distribution, and then from a Gaussian with the
appropriate width.  The numbers generated were values of $\chi-\psi$
for the positions and redshifts of the 71 sources in the sample with
$z\geq 0.3$.  In 1000 realizations of the data drawn from a
flat distribution, in only 7 trials was the significance of the
correlation greater than that in the actual data; this is comparable
to the 6 out of 1000 reported in \cite{nr}, and if reliable would
be evidence of the existence of a signal.  On the other hand, in
1000 realizations of the data drawn from the appropriate Gaussian
distribution, the artificial data was more strongly correlated with 
the hypothesized test function in 911 out of 1000 trials.  Even if
there were no signal at all in the data, we would expect the artificial
realizations to have a stronger correlation approximately $50\%$ of the 
time; the fact that our trials had better correlations over $90\%$ of
the time is due to the fact that the Gaussian slightly underestimates
the number of data points near $0^\circ$.  This result, however, 
vividly demonstrates our main point: the existence of a real enhancement
of $\chi-\psi$ near $90^\circ$ leads to a spuriously large correlation
coefficient if one uses the procedure described in \cite{nr}.  
When this enhancement, which is consistent with
conventional models of the sources, is taken into account, there is no
sign of an additional effect such as that in Eq.~\ref{deltachi}.

There is another way of quantifying our claim that a random distribution
centered around $90^\circ$ is a better fit to the data than the
correlation proposed in \cite{nr}.  Figure Five is a plot of 
$\chi-\psi$ as a function of $r\cos\theta$, where $\theta$ is defined
using the best-fit direction quoted in \cite{nr} and $\chi-\psi$ is
defined to be between $0^\circ$ and $180^\circ$.  
\begin{figure}
  \vskip -0.75cm
  \centerline{
  \psfig{figure=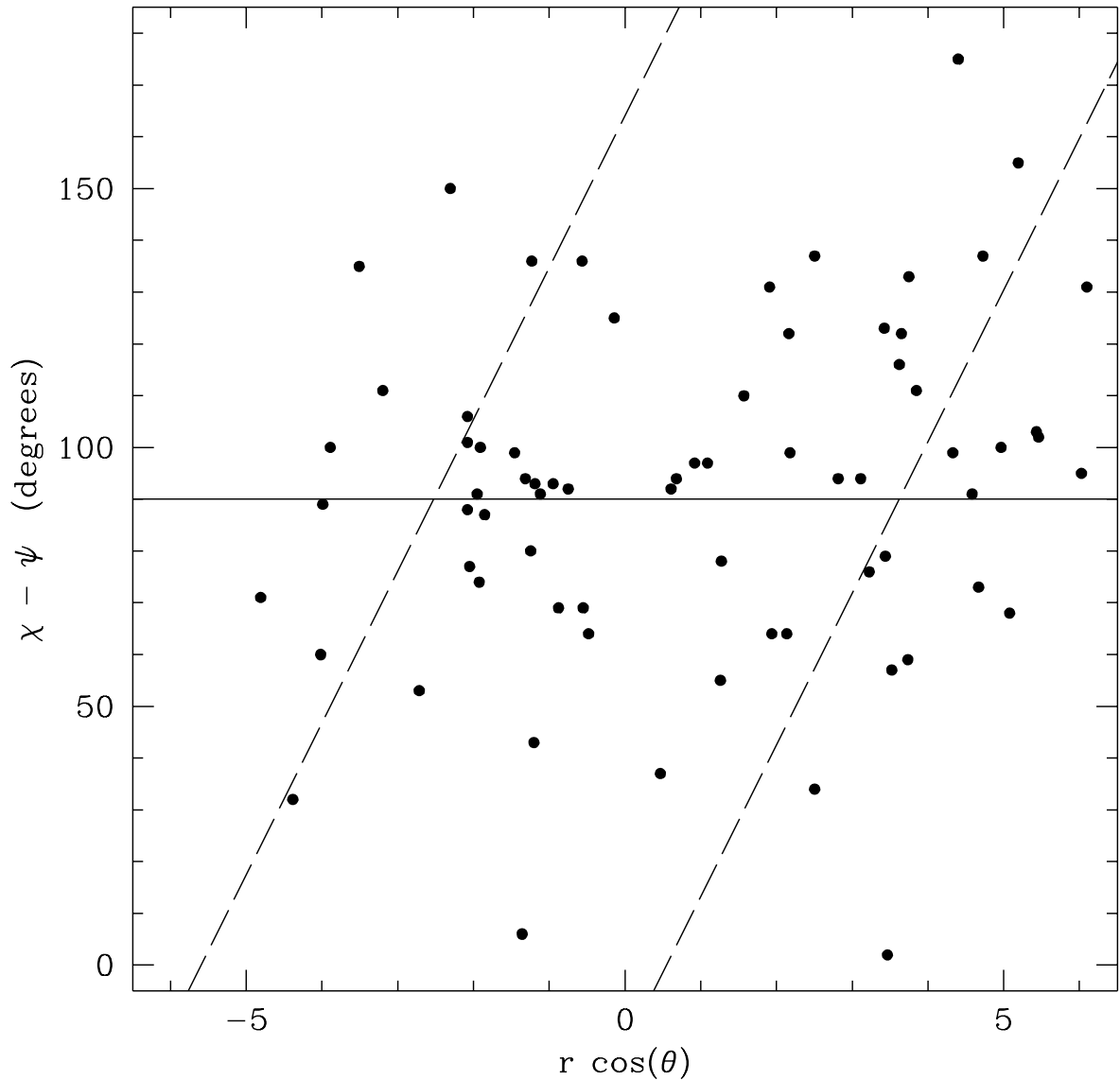,angle=0,height=16.5cm}}
  \vskip -0.1cm
  \caption{The difference between polarization and position angles
  as a function of $r\cos\theta$ for the best-fit direction of
  anisotropy proposed in \cite{nr}.  Angles of $180^\circ$ are
  to be identified with $0^\circ$; the data thus live on a cylinder.
  The solid line represents the predicted relationship in the absence
  of any signal, while the diagonal dashed line wrapping around the
  cylinder represents the model suggested in \cite{nr}.}
\end{figure}
We may think of
this graph as being defined on a cylinder, where $0^\circ$ is to
be identified with $180^\circ$.  With this in mind, we have plotted
two possible relationships, a solid horizontal line at $90^\circ$
and a dashed line at $(1/2)\Lambda_s^{-1} r\cos\theta+\delta$, where
we have measured the parameters $\Lambda_s$ and $\delta$ from
Figure 1(d) of \cite{nr}.  If the relationship claimed in \cite{nr}
is correct, the dashed line wrapping around the cylinder should be a
better fit to the data than the solid horizontal line.  This can be
measured by calculating
\begin{equation} 
  \chi^2 = \sum_i\left({{\Delta_i}\over{\sigma}}\right)^2\ ,
%  \label{}
\end{equation}
where we take the average error to be $\sigma = 33^\circ$, although
the precise value is irrelevant for purposes of comparison.  The
quantity $\Delta_i$, which represents the difference between the
predicted and measured value of $\chi-\psi$, is of course subject
to the $180^\circ$ ambiguity; however, we can resolve this ambiguity
optimistically for each point, by defining $-90^\circ < \Delta_i <
90^\circ$.  Using this procedure, we calculate that the best fit
proposed in \cite{nr} yields $\chi^2 = 161$, while the hypothesis
of no effect yields $\chi^2 = 69$.  Thus, the horizontal solid line
in Figure Five is a much better fit than the diagonal dashed lines.

Given that there is ample evidence that the intrinsic zero point
is centered on $\chi-\psi=90^\circ$, we may ask
how good a limit we can place on an effect such as that in 
Eq.~(\ref{deltachi}).  One approach to this problem is to define
$\chi-\psi$ to be between $0^\circ$ and $180^\circ$, and to assume
that the deviation from the intrinsic value is given by
$\Delta\chi = \chi-\psi - 90^\circ$.  It is then possible to do a
straightforward least-squares fit to Eq.~(\ref{deltachi}), with the
four components of $p_\mu$ as free parameters.  Using the data at
redshifts $z\geq 0.3$, the best-fit parameters obtained in this way are
\begin{equation} 
  \begin{array}{rcl}
  p_0 &=& (0.59 \pm 0.80)H_0 \\
  &=& (1.25 \pm 1.71)\times 10^{-42}h_0~{\rm GeV}\ ,\\
  |\vec p \,| &=& (1.13 \pm 1.40)H_0 \\
  &=& (2.41 \pm 2.99)\times 10^{-42}h_0~{\rm GeV}\ .\\
  \end{array}
%  \label{}
\end{equation}
(This procedure yields separate values for each of the three 
spacelike components $p_i$; since each value is consistent with
no preferred direction, it is more appropriate to quote the limit
on the magnitude $|\vec p \,|$.)
These values are consistent with $p_\mu=0$, and similar to the
limit on $p_0$ from \cite{cfj} quoted in Eq.~(\ref{limit}).

\section{Discussion}
\label{sec:disc}

After analyzing the data in a variety of ways, we are able to
conclude with confidence that there is no evidence for a chiral effect
on the propagation of photons from distant radio sources.  
Despite this negative result, there are still good reasons to
further pursue observations such as those examined in this paper.

\begin{figure}
  \vskip -0.75cm
  \centerline{
  \psfig{figure=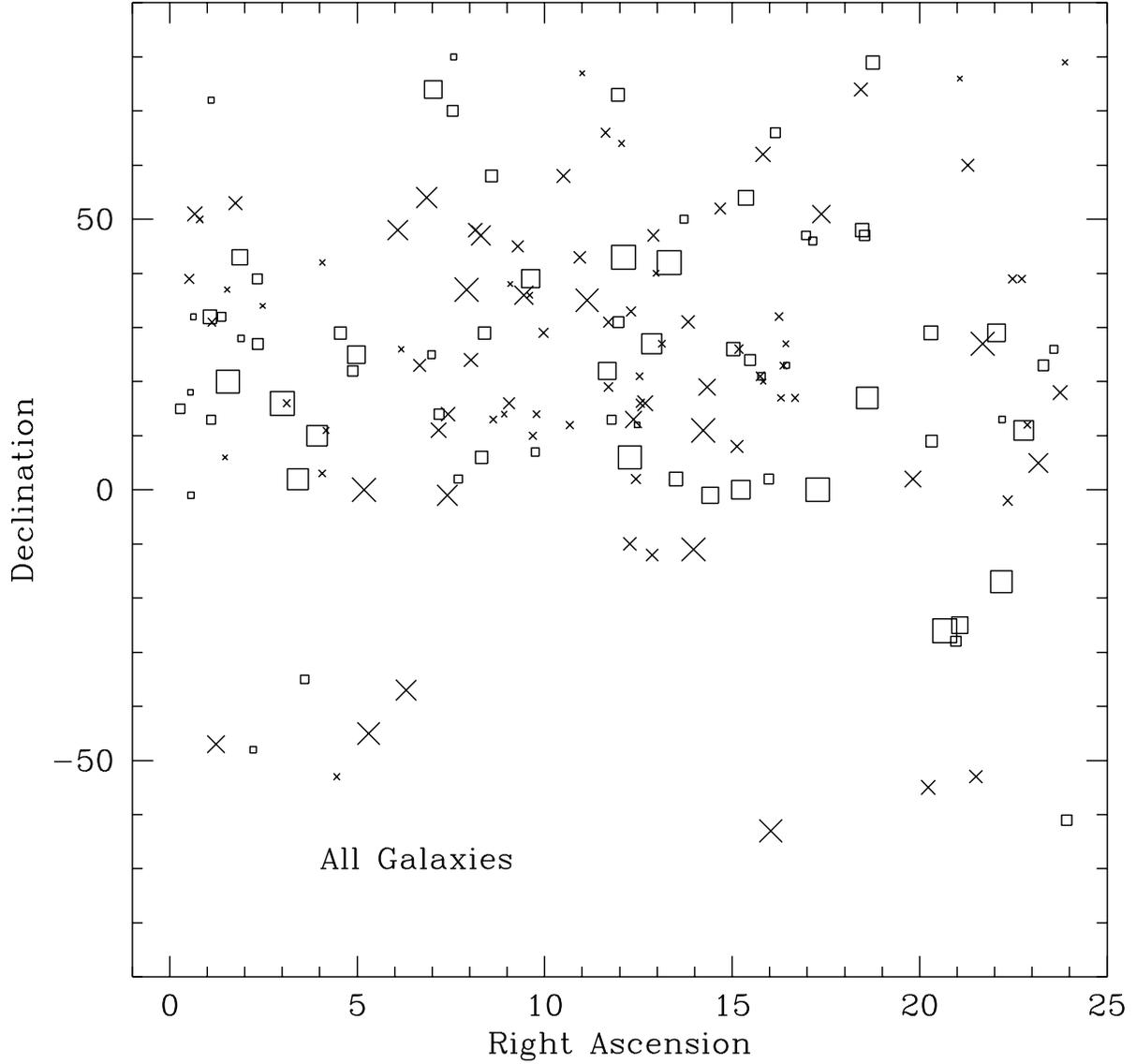,angle=0,height=16.5cm}}
  \vskip -0.1cm
  \caption{Positions of radio sources on the sky, including all redshifts.
  The symbols indicate deviations from $\chi-\psi=90^\circ$; squares
  are sources with $\chi-\psi<90^\circ$, and $\times$'s are sources
  with $\chi-\psi>90^\circ$.  The size of the symbol indicates the
  amount of deviation from $90^\circ$.}
\end{figure}

\begin{figure}
  \vskip -0.75cm
  \centerline{
  \psfig{figure=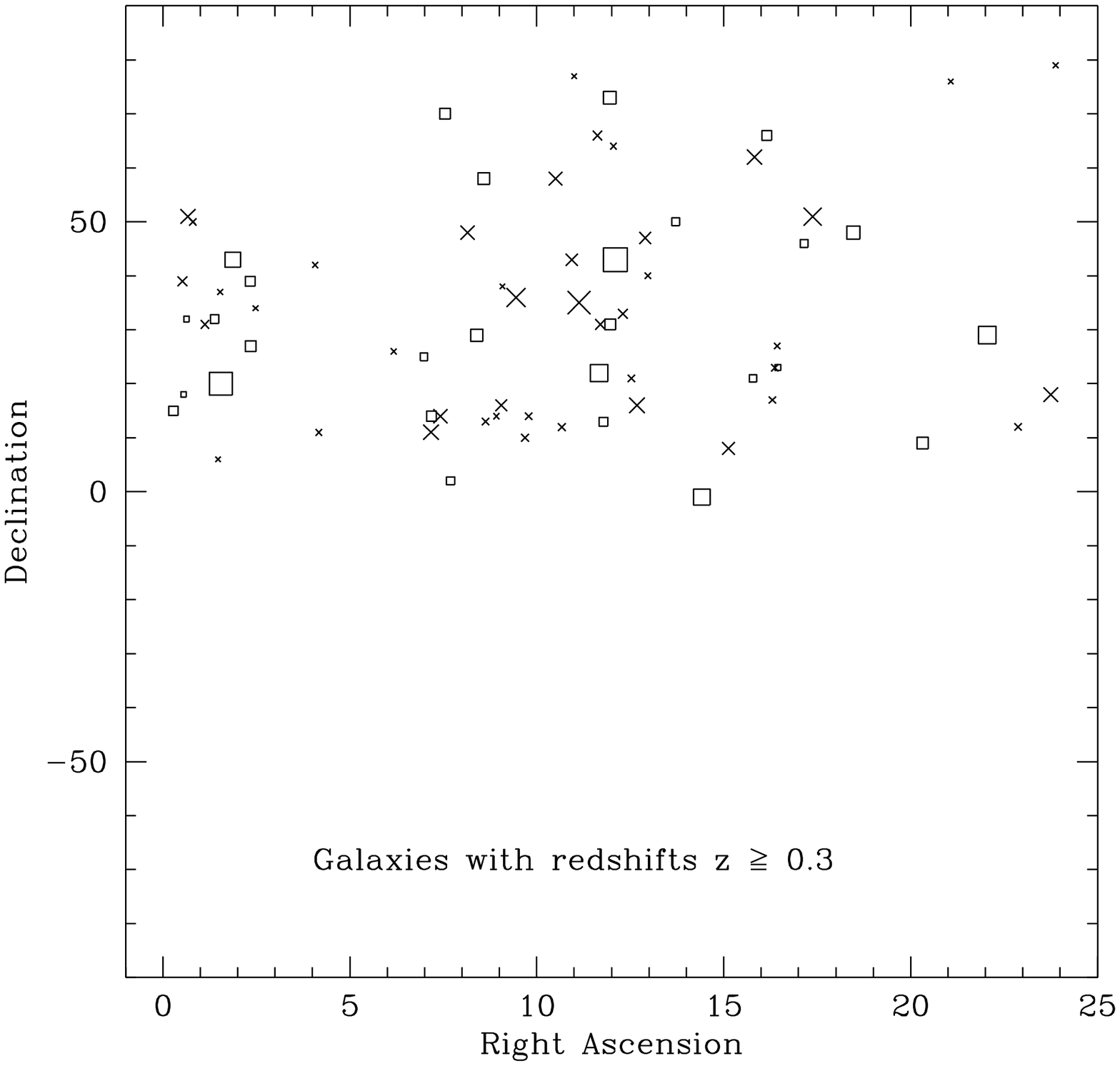,angle=0,height=16.5cm}}
  \vskip -0.1cm
  \caption{Same as Figure Six, including only galaxies with $z\geq 0.3$.}
\end{figure}

In Figures Six and Seven we have plotted the position of the
sources in the sky, indicated by symbols related to the deviation
of $\chi-\psi$ from $90^\circ$.  Figure Six includes all of the
galaxies, while Figure Seven is limited to the distant sources
with $z\geq 0.3$.  The squares represent sources with 
$\chi-\psi<90^\circ$, while the $\times$'s are sources
with $\chi-\psi>90^\circ$.  The size of the symbol is related
linearly to the deviation from $90^\circ$, although for clarity
there is an offset so that points with $\chi-\psi$ very close
to $90^\circ$ still have a nonzero size.  One conclusion to be
drawn immediately from these graphs is that there is a need for
additional data to be collected in the southern celestial
hemisphere, especially at high redshifts.  In the future, observations
of polarization of the cosmic microwave background may be the
best source of data for constraining phenomena such as these \cite{hs,sk}.

In characterizing the limits one can place on chiral effects, for
convenience we hypothesized a fixed four-vector $p_\mu$ which
would represent a violation of Lorentz invariance.  If an effect
were to be found, however, it is by no means necessary that such
a profound conclusion would have to be drawn.  A more plausible
hypothesis would be that of a very slowly-varying scalar field $\phi$
with a coupling as in (\ref{lag}); the application of the data
discussed in this paper to this possibility was examined in \cite{cf}.
Such a field could arise as an ultralight axion, with mass of order
the Hubble constant today (or less).  Interestingly, such axions may appear
naturally in the strongly coupled limit of heterotic string theory
\cite{bd1,bd2}.  Another possibility is the detection of 
axion-like cosmic strings; in the vicinity of such a string,
the polarization angle of two light rays passing on either side will 
undergo rotations in opposite directions \cite{hn}.  Although
there is no obvious sign of such a signal in Figures Six and Seven,
the importance of such a finding encourages us to continue the
search.

\eject
\section{Acknowledgments}

We would like to thank Ilana Harrus, 
Jack Hughes, Arthur Kosowsky, and Uros
Seljak for helpful conversations, and Borge Nodland for assistance
with the data.
This work was supported in part by the National Science Foundation
under grant PHY/94-07195 and NASA under grant NAGW-931.

\end{document}